\documentclass{amsart}

\usepackage{amsmath,amssymb,amsfonts,amscd,amsthm,amsopn,epsfig}  
\usepackage{amsthm}
\usepackage{amssymb}
\usepackage{latexsym,verbatim}
\usepackage{mathrsfs}
\usepackage{enumerate}

\usepackage{tikz}\usetikzlibrary{backgrounds,matrix,arrows,shapes,decorations.markings}
\newcommand{\btp}{\begin{tikzpicture}[baseline=-5pt,scale=0.25,line width=0.7pt]}
\newcommand{\etp}{\end{tikzpicture}}

\pgfdeclarelayer{background}
\pgfdeclarelayer{foreground}
\pgfsetlayers{background,main,foreground}

\usepackage{hyperref}
\hypersetup{
    citecolor=blue,        
}

\usepackage{color}

\newtheorem{prop}{Proposition}[section]

\newtheorem{rmk}{Remark}[section]

\numberwithin{equation}{section}

\def\V={{{\bf\rm{V}}}}

\def\beq{\begin{equation}}
\def\eeq{\end{equation}}
\def\bea{\begin{eqnarray}}
\def\eea{\end{eqnarray}}
\def\ba{\begin{array}}
\def\ea{\end{array}}

\def\lt{\left}
\def\rt{\right}

\def\beq{\begin{equation}}
\def\eeq{\end{equation}}
\def\ben{\begin{eqnarray}}
\def\een{\end{eqnarray}}
\def\ba{\begin{array}}
\def\ea{\end{array}}

\def\lt{\left}
\def\rt{\right}

\newcommand\cW{{\mathcal W}} 
\newcommand\cH{{\mathcal H}}

\newcommand\CC{\mathbb C}

\begin{document}

\title{Modified algebraic Bethe ansatz for XXZ chain on the segment\\  - III - proof}
\author{J.~Avan}
\address{Laboratoire de Physique Th\'eorique
et Mod\'elisation (CNRS UMR 8089), Universit\'e de Cergy-Pontoise, F-95302 Cergy-Pontoise, France }
\email{jean.avan@u-cergy.fr}
\author{S.~Belliard}
\email{samuel.belliard@u-cergy.fr}
\author{N.~GrosJean}
\email{nicolas.grosjean@u-cergy.fr}
\author{R.A.~Pimenta}
\address{Departamento de F\'{\i}sica, Universidade Federal de S\~ao Carlos,
Caixa Postal 676, CEP 13569-905, S\~ao Carlos, Brasil}
\address{Physics Department, University of Miami, P.O. Box 248046, FL 33124, Coral Gables, USA}
\email{pimenta@df.ufscar.br}

\begin{abstract}
In this paper, we prove the off-shell equation satisfied by the transfer matrix associated with the 
XXZ spin-$\frac12$ chain on the segment with two generic integrable boundaries acting on the Bethe vector. The essential step is to prove that  the expression of the
action of a modified creation operator on the Bethe vector has an off-shell structure which results in an inhomogeneous term
in the eigenvalues and Bethe equations of the corresponding transfer matrix.
\end{abstract}

\maketitle

\vskip -0.2cm




\section{Introduction}


In the last decade, the study of the spectral problems associated with quantum integrable models which are not invariant by $U(1)$ symmetry has been an active research field. 
The prototype model is the XXZ spin-$\frac{1}{2}$ chain on the segment with general boundary conditions, see \cite{Bel14,BelPim14} for the Hamiltonian in our notations. 
For non-diagonal boundaries, although the model is known to be integrable, the Bethe ansatz techniques, and in particular the algebraic Bethe ansatz (ABA) \cite{SFT,Skl88}, could not be directly applied due to the lack of a reference state for the transfer matrix\footnote{For models with $U(1)$  symmetry and $q$ not root of unit, the reference state corresponds to the highest weight vector of the finite dimensional representation of the underlying algebra.}. There were many attempts to overcome this problem: by introducing constraints between the left and the right boundary parameters \cite{gauge,YanZ07,TQ,TQ2}, by considering specific non-diagonal integrable boundaries\footnote{In a way such that a reference state for the transfer matrix can still be identified.} and introducing a new structure for the Bethe vector \cite{BCR12,PL13}, or by developing alternative methods to the Bethe ansatz  \cite{BK,Gal08,LP14}, including the {\it  separation of variables} (SoV) \cite{niccoli2,FKN13} method. More historic details can be found in  \cite{Bel14}.

\vspace{0.1cm}

Recently, an essential step to construct the solution of quantum integrable models without $U(1)$ symmetry by means of the Bethe ansatz was accomplished by the introduction of the so-called {\it off-diagonal Bethe ansatz} (ODBA)  \cite{CYSW1,CYSW3}.  The key point is the use, as an ansatz to characterize the spectrum of such models, of a modified T-Q Baxter equation which exhibits an additional inhomogeneous term.

\vspace{0.1cm}

These results brought the questions: what are the Bethe vector associated with the new eigenvalues and, in addition, can the new inhomogeneous term be obtained solely from an ABA perspective?

\vspace{0.1cm}

The first progress arose in \cite{BC13}, where the Bethe vector of the XXX spin-$\frac{1}{2}$ chain on the segment  were conjectured from the modified T-Q Baxter equation \cite{CYSW3}. In  \cite{Bel14, BelPim14} the origin of the additional inhomogeneous term was clarified by means of a modified algebraic Bethe ansatz (MABA). In the MABA we drop the quest of the reference state and just focus on the construction of a highest weight vector for the underlying reflection algebra.  
In this framework, the additional term is understood as the action of a modified creation operator on the off-shell Bethe vector\footnote{A Bethe vector is said to be off-shell if it depends on arbitrary parameters and on-shell if these parameters satisfy the Bethe equations.}. Such off-shell action was proved for the totally asymmetric simple exclusion process (TASEP) model (a degenerate case of the XXZ spin-$\frac{1}{2}$ chain on the segment with non-diagonal boundaries) in \cite{Cra14}. This proof uses the so-called F-basis \cite{MS96} to analyze the pole structure of the creation operator. Let us also mention that a proof of the Bethe vector expression, solely for the on-shell case, was proposed from the ODBA for the XXX,  in \cite{CYSW4}, and XXZ, in \cite{CYSW5}, spin-$\frac{1}{2}$ chains on the segment. In this case, the canonical SoV-basis \cite{FKN13} was used.

\vspace{0.1cm}

Here we prove the validity of MABA approach for the transfer matrix of the finite XXZ spin-$\frac{1}{2}$ chain on the segment with two generic boundaries.
The expression of the off-shell Bethe vector of this model were conjectured in \cite{BelPim14}. Let us recall that the construction
of the Bethe vector follows the steps:
\begin{itemize}
\item Find the highest weight representation for the operators that describe the model through the generalized quantum inverse scattering method \cite{Skl88}. This was done in \cite{gauge} from a local gauge transformation. This is possible if the dynamical right reflection matrix is diagonal or upper-triangular.
\item Use the gauge transformation to map the transfer matrix to a modified lower triangular form \cite{BelPim14} or to a modified diagonal form \cite{BC13}. This step introduces the modified operator used to construct the Bethe vector (which is built from applying successively the modified operator on the highest weight vector).
\item Apply the transfer matrix on the Bethe vector. Use the commutation relations and the action of the modified operators on the highest weight representation to obtain the off-shell action of the transfer matrix. This action has a new term with an additional modified creation operator acting on the Bethe vector.
\item Show that the new term has a wanted-unwanted structure. From direct resolution of chains with small length ($N=1,2$), we notice that this is possible if the number of creation operators equals the length of the chain. We observe that these wanted-unwanted terms give the inhomogeneous part in the eigenvalues and in the Bethe equations of the transfer matrix introduced in the context of the ODBA. This step was performed in  \cite{BelPim14}.
\end{itemize}

In this paper, we prove the explicit form of the off-shell action of the modified creation operator on the off-shell Bethe vector using the canonical SoV-basis \cite{FKN13}.
As an additional result, we make the explicit link between the MABA and the SoV  \cite{FKN13,niccoli2}  and  ODBA  \cite{CYSW3,CYSW4,CYSW5} methods.

\vspace{0.1cm}

The paper is organized as follows. In section \ref{sectQISM}, we recall the definition    
of the transfer matrix of XXZ spin-$\frac{1}{2}$ chain on the segment in the generalized quantum inverse scattering framework as well as the gauge transformation that yields the dynamical operators.  In section \ref{sectRep}, we recall the highest weight representation of the dynamical operators and introduce the modified dynamical operators. In section \ref{sectMABA}, we implement the MABA and in section \ref{sectProof} we prove the action formula of the modified creation operator on the off-shell Bethe vector. The connection with the ODBA and SoV methods is given in section \ref{sectODBA-SoV}. We conclude in section \ref{sectCon}.  

\section{Generalized quantum inverse scattering and gauge tranformation \label{sectQISM}}

The transfer matrix of the finite XXZ spin-$\frac12$ chain on the segment is
constructed following the generalized quantum inverse scattering
framework \cite{Skl88}. Here we recall only the essential features, more details in our notations can be found in \cite{Bel14,BelPim14}.
From the trigonometric $R-$matrix,  which acts on $V_a\otimes V_b$\footnote{In this work, we only use two-dimensional complex vector spaces, {\it i.e.} $V=\CC^2$.},
\ben\label{R}
\qquad R_{ab}(u)=\lt(\begin{array}{cccc}b(qu)&0&0&0\\0&b(u)&1&0\\
0&1&b(u)&0\\0&0&0&b(qu)\end{array}\rt),\quad
b(u)=\frac{u-u^{-1}}{q-q^{-1}},
\een
and from the $K-$matrices \cite{dVGR1}, which act on $V$,
\ben\label{Km}
&&K^-(u)=\lt(\begin{array}{cc}k^-(u)&\tau^2 \,c(u)\\
\tilde \tau^2 \,c(u)&k^-(u^{-1})\end{array}\rt),
\quad k^-(u) =\nu_-u+\nu_+u^{-1},\quad  c(u)=u^2-u^{-2},\\
\label{Kp}
&&K^+(u)=
\lt(\begin{array}{cc}
k^+(qu)&\tilde \kappa^2 \,c(qu)\\
\kappa^2 \,c(qu)&k^+(q^{-1}u^{-1})
\end{array}\rt),\quad k^+(u) =\epsilon_+u+\epsilon_-u^{-1},
\een
where $\{\epsilon_\pm,\kappa, \tilde \kappa \}$
and $\{\nu_\pm,\tau,\tilde \tau \}$ are generic parameters, one constructs the transfer matrix, that acts on the Hilbert space $\cH=\otimes_{i=1}^N V_i$,
\ben\label{tr}
\qquad t(u)=tr_a(K^+_a(u)K_a(u)),
\een
where
\ben\label{K}
&&K_a(u)=R_{a1}(u/v_1)\dots R_{aN}(u/v_N)K_a^-(u)R_{aN}(uv_N)\dots R_{a1}(uv_1),\\
&&\qquad\quad=\left(\begin{array}{cc}
       \mathscr{A}(u) & \mathscr{B}(u)\\
       \mathscr{C}(u) & \mathscr{D}(u)+\frac{1}{b(qu^2)}\mathscr{A}(u)
      \end{array}
\right)_a\label{KO},
\een 
is the double row monodromy matrix that acts on $V_a\otimes \cH$.
This formalism allows one to reformulate the problem in the operator language. The transfer matrix is given by,
\ben
t(u)=
\phi(u)k^+(u)\mathscr{A}(u) +
k^+(q^{-1}u^{-1})\mathscr{D}(u)+c(qu)
\Big(\kappa^2 \,\mathscr{B}(u)+\tilde\kappa^2 \,\mathscr{C}(u)\Big),
\een
with
\ben\label{defphi}
\phi(u)=\frac{b(q^2u^2)}{b(q u^2)}.
\een

\vspace{0.1cm}

An important point in the ABA framework is the existence of a highest
weight vector which is annihilated by the operator $\mathscr{C}(u)$
and which is an eigenstate of the operators $\mathscr{A}(u)$ and $\mathscr{D}(u)$.
This vector exists when we have an upper triangular $K^-$ matrix. However,
for generic $K^-$ matrix, one needs to use the gauge transformation
\cite{baxter,gauge} in order to define new double-row operators that are dynamical, {\it i.e.}
they depend of an additional parameter said dynamical, and to be able to
construct such highest weight vector. In addition, the gauge transformation allow us to eliminate the dynamical creation and/or annihilation operators
in the transfer matrix expression. The building blocks of the gauge transformation are the covariant vectors,
\bea\label{coVec}
&& |X(u,m)\rangle=\left(\begin{array}{c}
     \alpha q^{-m} u^{-1} \\
       1
      \end{array}
\right),\quad
|Y(u,m)\rangle=\left(\begin{array}{c}
     \beta q^{m} u^{-1} \\
       1
      \end{array}
\right),\een
and the contravariant vectors,
\ben
\label{conVec}
&& \langle \tilde X(u,m)|=\frac{q u}{\gamma_{m-1}}\left(\begin{array}{cc}
-1,& \alpha q^{-m} u^{-1}
\end{array}
\right),\quad
 \langle \tilde Y(u,m)|=\frac{q u}{\gamma_{m+1}}\left(\begin{array}{cc}
1 ,&
      - \beta q^{m} u^{-1} 
      \end{array}
\right),
\eea
where $\alpha$ and $\beta$ are generic complex parameters,
$m$ is an integer and $\gamma(u,m)=\alpha q^{-m} u-\beta q^{m} u^{-1}$ with $ \gamma(1,m)=\gamma_m$.
They satisfy the shifted scalar products,
\bea\label{Scal-prod}
&& \langle\tilde X(u,m)|X(u,m)\rangle= \langle\tilde Y(u,m)|Y(u,m)\rangle=0, \\
&& \langle \tilde X(u,m+1)|Y(u,m-1)\rangle= \langle\tilde Y(u,m-1)|X(u,m+1) \rangle=1,
\eea
as well as the shifted closure relation,
\bea\label{close}
&& |Y(u,m-1)\rangle\langle\tilde X(u,m+1)|+|X(u,m+1)\rangle\langle \tilde Y(u,m-1)|= \left(\begin{array}{cc}
       1 & 0\\
      0 & 1      \end{array}\right).
\eea
From these vectors we can introduce the dynamical operators,
\bea\label{odyn}
&&\mathscr{C}(u,m) =\langle\tilde X(u,m)|K_a(u)|X(u^{-1},m)\rangle,\\
&&\mathscr{B}(u,m)=\langle \tilde Y(u,m)|K_a(u)|Y(u^{-1},m)\rangle,\\
&&\mathscr{A}(u,m)=\langle\tilde Y(u,m-2)|K_a(u)|X(u^{-1},m)\rangle, \\
&&\mathscr{D}(u,m)=\frac{\gamma_{m+1}}{\gamma_m}\langle\tilde X(u,m+2)|K_a(u)|Y(u^{-1},m)\rangle-
\frac{\gamma(u^{-2},m+1)}{b(qu^2)\gamma_m} \mathscr{A}(u,m).
\eea
such that the transfer matrix can be written in the following way,
\ben\label{t}
t(u)=t_d(u,m)+
u^{-1}c( q u)\Big( \zeta_{m}\mathscr{B}(u,m)-\tilde \zeta_m \mathscr{C}(u,m)\Big),
\een
with
\ben
&& \zeta_m =\frac{\kappa^2}{\gamma_m}
\Big( \alpha q^{-m-1} +i\frac{\tilde\kappa\xi}{ \kappa\tilde \xi }\Big)
\Big( \alpha q^{-m-1} +i \frac{\tilde\kappa\tilde\xi}{\kappa \xi }\Big),
\quad \tilde \zeta_m =\frac{\kappa^2}{\gamma_m}
\Big( \beta q^{m-1} +i  \frac{\tilde \kappa \xi}{ \kappa \tilde \xi }\Big)
\Big( \beta q^{m-1} +i \frac{ \tilde\kappa \tilde\xi}{  \kappa \xi }\Big),
\een
and the new parametrization,
\ben\label{NpKm}
&&\epsilon_-=i\tilde\kappa\kappa\big(\xi/\tilde \xi+\tilde\xi/ \xi\big),
\quad \epsilon_+=i\tilde\kappa\kappa\big(\xi \tilde \xi+1/(\tilde\xi \xi)\big).
\een
The term $t_d(u,m)$, which is the diagonal part of the transfer matrix, is given by
\ben\label{td}
\qquad\quad t_d(u,m)=\tilde a(u,m)\mathscr{A}(u,m)+\tilde d(u,m)\mathscr{D}(u,m),
\een
where
\ben
\quad \tilde a(u,m)= \tilde a (u)-\delta_m\phi(q^{-1}u^{-1})u^{-1}c( q u),\quad \tilde d(u,m)= \tilde d (u)+u^{-1}c( q u)\delta_m
\een
with
\ben
 \tilde a (u)=u^{-1}\phi(u)\tilde k^+(u), \quad \tilde d (u)=u^{-1}\tilde k^+(q^{-1}u^{-1}),
\\
\tilde k^+(u)=
i \tilde\kappa\kappa   ( \tilde \xi u +\tilde \xi^{-1} u^{-1})(\xi^{-1}u + \xi u^{-1}),
\eea
and
\ben
\delta_m=
\frac{\kappa^2}{\gamma_{m+1}} \Big(\alpha q^{-m-1}
+i \frac{\tilde \kappa  \xi}{\kappa \tilde \xi }\Big)
\Big(\beta q^{m+1}+i \frac{\tilde\kappa\tilde \xi}{ \kappa\xi } \Big).
\een


Let us introduce finally a bookkeeping notation.
For generic gauge parameters $\{\alpha,\beta\}$ the dynamical operators are denoted by,
\ben
\{ \mathscr{A}(u,m),\mathscr{B}(u,m),\mathscr{C}(u,m),\mathscr{D}(u,m) 
\}. \nonumber
\een
If we consider some fixed  gauge parameters $\{  \alpha_x,\beta_x\}$ we denote the corresponding fixed dynamical operators with indices, namely,
\ben
\{ \mathscr{A}_x(u,m),\mathscr{B}_x(u,m),\mathscr{C}_x(u,m),\mathscr{D}_x(u,m) \nonumber
\}. 
\een


\section{Representation theory and modified operators \label{sectRep}}


In order to consider the diagonalization of the transfer matrix in the framework of the generalized quantum inverse scattering method, we want first to introduce a suitable basis to construct the Bethe vector. In particular, since we want a highest weight representation, it can be built for the dynamical operators with the choice of gauge parameters
\ben\label{alphahw}
\alpha\Rightarrow\alpha_{dr}= i q^{m_0+N} \frac{\tau  \mu}{\tilde\tau \tilde \mu}, \quad \beta\Rightarrow\beta_{dr}=i q^{-m_0-N} \frac{\tau\tilde \mu}{\tilde\tau  \mu},
\een
which corresponds to bring the dynamical right $K^-$-matrix to a diagonal form, see \cite{gauge} for details. As a consequence,
it allows us to construct the highest weight vector,
\bea\label{Dhwv}
|\Omega_{m_0}^N\rangle_{dr}=\otimes_{i=1}^N|X(v_i,m_0+i)\rangle_{dr}.
\eea
such that the actions of the dynamical operators at the point $m=m_0$ are given by,
\bea\label{actDhwv}
&&\mathscr{A}_{dr}(u,m_0)|\Omega_{m_0}^N\rangle_{dr}=u\, \tilde k^-(u) \Lambda(u)|\Omega_{m_0}^N\rangle_{dr},\\
\label{actDhwv1}&&\mathscr{D}_{dr}(u,m_0)|\Omega_{m_0}^N\rangle_{dr}=u \,\phi(q^{-1}u^{-1})\tilde k^-(q^{-1}u^{-1})
\Lambda(q^{-1}u^{-1})|\Omega_{m_0}^N\rangle_{dr},\\
\label{actDhwv2}&&\mathscr{C}_{dr}(u,m_0)|\Omega_{m_0}^N\rangle_{dr}=0,
\eea
where
\ben\label{defLAM}
 \Lambda(u) = \prod_{i=1}^{N}b(qu/v_i)b(quv_i),
\een
and
\ben
\tilde k^-(u)=
i \tilde\tau\tau ( \mu u +\mu^{-1} u^{-1})(\tilde\mu^{-1}u + \tilde\mu u^{-1}),
\een
with the new parametrization
\ben
\nu_-=i\tilde\tau\tau\big(\mu/\tilde \mu+\tilde \mu/\mu\big),
 \quad \nu_+=i\tilde\tau\tau\big(\mu\tilde\mu+1/( \mu\tilde \mu)\big).
\een
Moreover the dynamical operator $\mathscr{B}_{dr}$  is dynamically nilpotent of order $N+1$,
\ben
\mathscr{B}_{dr}(u,m_0+2N)\mathscr{B}_{dr}(u_1,m_0+2(N-1))\dots \mathscr{B}_{dr}(u_N,m_0)=0,
\een  
where $\bar u=\{u_{1},\dots,u_{N}\}$ is a set of $N$ generic parameters.

Let us now consider the string of $M$ operators acting on the highest weight vector,
\ben\label{vecBase}
\mathscr{B}_{dr}(u_{j_1},m_0+2(M-1))\dots \mathscr{B}_{dr}(u_{j_M},m_0)|\Omega_{m_0}^N\rangle_{dr}
\een
where $M\in\{0,\dots,N\}$  and $\{u_{j_1},\dots,u_{j_M}\}$ is some subset of cardinality $M$ of $\bar u$. Since the vectors (\ref{vecBase}) are independent for generic $\bar u$, 
from these vectors and for fixed $M$ we generate a subspace $\cW_M^{m_0}$ of dimension  $\Big(\!\!\begin{array}{c}  N \\  M \end{array}\!\!\Big)=\frac{N!}{M!(N-M)!}$
 in the Hilbert space. The highest weight representation induces the decomposition
 \ben
 \cH=\bigoplus_{M=0}^N\cW_M^{m_0}
 \een
 of the Hilbert space. Indeed the total dimension of the sum on the r.h.s. is  
 \ben
\sum_{M=0}^N\Big(\!\!\begin{array}{c}  N \\  M \end{array}\!\!\Big)=2^N.
\een

\vspace{0.1cm}

Thus, in order to construct the Bethe vector of the model, we are looking for a linear combination of the vectors (\ref{vecBase}).
However, fixing the full family of coefficients is a hard task. It was performed for small $N$ in the case of the XXX chain on the segment \cite{BC13}, by using the {\it a priori}  knowledge of the eigenvalues. It appears that the Bethe vector factorizes in terms of a modified creation operator. This modified creation operator and the associated family of other modified operators are the key step to implement the MABA without the knowledge of the eigenvalues. These operators are linear combinations of the ones used to construct the highest weight representation. This modified family of operators reduces the transfer matrix to a modified diagonal form (see below) and the commutation relations are  similar to the diagonal boundary case. Thus, a part of the calculation of the MABA is similar to the usual ABA and reduces to already known cases by simple limit.
Indeed, for the choice of the gauge parameters,

\ben\label{gaugelowup}
\alpha\Rightarrow\alpha_{dl}(M)=-iq^{1+m_0+2M}\frac{\xi \tilde\kappa}{\tilde\xi\kappa},\quad\quad\quad\quad \beta\Rightarrow\beta_{dl}(M)=-iq^{1-m_0-2M}\frac{\tilde\xi \tilde\kappa}{\xi\kappa},
\een 
the dynamical transfer matrix (\ref{td}) at the point $m_0+2M$ takes a modified diagonal form, {\it i.e. it depends only in $\mathscr{A}$ and $\mathscr{D}$ operators},  namely
\ben\label{tm}
&& t(u)=t_{dl}(u,m_0+2M)=\tilde a(u)\mathscr{A}_{dl}(u,m_0+2M)+\tilde d(u)\mathscr{D}_{dl}(u,m_0+2M).
\een
Thus it becomes natural to consider $\mathscr{B}_{dl}(u,m)$ as a modified creation operator that we will use to construct the Bethe vector. In the following,  to apply the ABA, we will need the commutation relations
\begin{eqnarray}
&&\mathscr{B}_{dl}(u,m+2)\mathscr{B}_{dl}(v,m) = \mathscr{B}_{dl}(v,m+2)\mathscr{B}_{dl}(u,m),\label{comBdBd} \\
&&\nonumber\mathscr{A}_{dl}(u,m+2)\mathscr{B}_{dl}(v,m)=f(u,v)\mathscr{B}_{dl}(v,m)\mathscr{A}_{dl}(u,m) +
\\&&\qquad \qquad g_{dl}(u,v,m)\mathscr{B}_{dl}(u,m)\mathscr{A}_{dl}(v,m) + w_{dl}(u,v,m)\mathscr{B}_{dl}(u,m)\mathscr{D}_{dl}(v,m),\label{comAdBd}  \\
&&\nonumber \mathscr{D}_{dl}(u,m+2)\mathscr{B}_{dl}(v,m)= h(u,v)\mathscr{B}_{dl}(v,m) \mathscr{D}_{dl}(u,m) 
\\&&\qquad \qquad+k_{dl}(u,v,m)\mathscr{B}_{dl}(u,m)\mathscr{D}_{dl}(v,m)+ n_{dl}(u,v,m)\mathscr{B}_{dl}(u,m)\mathscr{A}_{dl}(v,m),\label{comDdBd}
\end{eqnarray}
obtained from the method described in \cite{gauge} and with functions defined in the appendix \ref{App:Func}. See also \cite{BelPim14} for details.
 
An important fact is that the operators of the two families are related by linear relations. In particular we have 
\ben
&&\mathscr{A}_{dl}(u,{p})
=\mathscr{A}_{dr}(u,{m})+c(p,m)\mathscr{C}_{dr}(u,{m})+b(p,m)\mathscr{B}_{dl}(u,{p}-2)\\
&&\mathscr{D}_{dl}(u,{p})
=\mathscr{D}_{dr}(u,{m})-\phi(u)c(p,m)\mathscr{C}_{dr}(u,{m})-\phi(u)b(p,m)\mathscr{B}_{dl}(u,{p}-2)
\een
with
\ben
b(p,m)=\frac{\alpha_{dr}-q^{m-p}\alpha_{dl}}{\alpha_{dr}-q^{m+p-2}\beta_{dl}}, 
\quad c(p,m)=\frac{q^{m-p}\beta_{dr}-\beta_{dl}}{q^{2-m-p}\alpha_{dr}-\beta_{dl}}.
\een 
It allows to find the action of the operators of the modified transfer matrix (\ref{td}) on the highest  weight vector (\ref{Dhwv}), namely
 \ben\label{ODactA}
&&\mathscr{A}_{dl}(u,m_0)|\Omega_{m_0}^N\rangle_{dr}=
u\, \tilde k^-(u) \Lambda(u)|\Omega_{m_0}^N\rangle_{dr}
+\eta_{m_0}\mathscr{B}_{dl}(u,m_0-2)|\Omega_{m_0}^N\rangle_{dr},\\
\label{ODactD}
&&\mathscr{D}_{dl}(u,m_0)|\Omega_{m_0}^N\rangle_{dr}=
u \,\phi(q^{-1}u^{-1})\tilde k^-(q^{-1}u^{-1})
\Lambda(q^{-1}u^{-1})|\Omega_{m_0}^N\rangle_{dr} \\
&&\qquad \qquad \qquad \qquad \qquad \qquad  - \phi(u)\eta_{m_0}\mathscr{B}_{dl}(u,m_0-2)|\Omega_{m_0}^N\rangle_{dr},\nonumber
\een
with
\ben
\eta_{m_0}=b(m_0,m_0)=\frac{\alpha_{dr}-\alpha_{dl}}{\alpha_{dr}-\beta_{dl}q^{2m_0-2}}.
\een
\begin{rmk}
These actions have an off-diagonal contribution characteristic of the MABA already pointed out in \cite{BC13,Bel14,BelPim14}. 
\end{rmk}


\section{MABA \label{sectMABA}}


From the modified dynamical $\mathscr{B}_{dl}$ operator, used as creation operator, we construct the string of $M$ operators 
\ben\label{SB}
B_{dl}(\bar u,m,M)=\mathscr{B}_{dl}(u_1,m+2(M-1))\dots \mathscr{B}_{dl}(u_M,m),
\een
and act with it on the highest weight vector (\ref{Dhwv}). It yields the vector,
\ben\label{Psi}
|\Psi_{m_0}^{M}(\bar u)\rangle=B_{dl}(\bar u,m_0,M)|\Omega^N_{m_0}\rangle_{dr}.
\een
Using the commutation relations (\ref{comBdBd},\ref{comAdBd},\ref{comDdBd}),
the off-diagonal actions on the highest weight vector (\ref{ODactA},\ref{ODactD}), and
the functional relations  (\ref{FR1},\ref{FR2}) to simplify the term with the additional $\mathscr{B}_{dl}$ operator,
we obtain from the usual algebraic Bethe ansatz arguments, see \cite{BelPim14} for more details, the off-shell action for the modified transfer matrix,
\ben\label{tdPsi}
&&t_{dl}(u,m_0+2M)|\Psi_{m_0}^{M}(\bar u)\rangle= \Lambda^M_{d}(u,\bar u)|\Psi_{{m_0}}^{M}(\bar u)\rangle+\sum_{i=1}^M\tilde F(u,u_i)E^M_{d}(u_i,\bar u_i)|\Psi_{{m_0}}^{M}(\{u,\bar u_i\})\rangle\\
&&\qquad\qquad\qquad\qquad \nonumber +\hat \eta_{m_0}u^{-1}c(qu)B_{dl}(\{u,\bar u\},m_0-2,M+1)|\Omega^N_{m_0}\rangle_{dr},
\een
with
\bea
&& \label{Lamdg}\Lambda_{d}^M(u,\bar u)=\phi(u)\tilde k^+(u)\tilde k^-(u)\Lambda(u) f(u,\bar u)+\phi(q^{-1}u^{-1})\tilde k^+(q^{-1}u^{-1})\tilde k^-(q^{-1}u^{-1})\Lambda(q^{-1}u^{-1}) h(u,\bar u),
\\
&&\label{Edg}E_{d}^M(u_i,\bar u_i) =\phi(q^{-1}u^{-1}_i)\phi(u_i)\Big(\tilde k^+(u_i)\tilde k^-(u_i)\Lambda(u_i)f(u_i,\bar u_i)\label{Egd}\\
&&\qquad  \qquad  \qquad \qquad  \qquad
-\tilde k^+(q^{-1}u_i^{-1})\tilde k^-(q^{-1}u_i^{-1})\Lambda(q^{-1}u_i^{-1}) h(u_i,\bar u_i)\Big),\nonumber \\
&& \hat \eta_{m_0}=q^{-1} \kappa^2\gamma^{dl}_{m_0-1}\eta_{m_0}\label{etahatdef}.
\een
and where we denote the set $\bar u= \{u_1, \dots, u_M\}$ with cardinality $M$, the set
$\bar u_i =\{u_1,u_2,\dots,u_{i-1},u_{i+1},\dots,u_M\}$,
where the element $u_i$ is removed, and the set $\{u,\bar u_i\}$ as the union of $u$ and $\bar u_i$.

\begin{rmk}
From the requirement $\hat \eta_{m_0}=0$ we recover the result of \cite{gauge} which implies constraints between left and right boundary parameters. See \cite{Bel14,BelPim14} for more details about the limit cases.
\end{rmk}

The new feature, characteristic of the MABA, is the presence of the string of $M+1$ creation operators in the off-shell equation (\ref{tdPsi}). This term
can be handled according to the following proposition,

\begin{prop}\label{P1} For $M=N$, the length of the spin chain, the string of $N+1$ modified creation operators
$ \mathscr{B}_{dl}$ acting on the highest  weight vector (\ref{Dhwv}) 
has an off-shell structure given by,
\ben\label{offshellB}
&&\hat \eta_{m_0}u^{-1}c(qu)B_{dl}(\{u,\bar u\},m_0-2,N+1)|\Omega^N_{m_0}\rangle_{dr}
=\Lambda^N_{g}(u,\bar u)|\Psi_{m_0}^N(\bar u)\rangle\nonumber\\
&&\qquad\qquad\qquad\qquad\qquad\qquad\qquad
+
\sum_{i=1}^N\tilde{F}(u,u_i)E^N_{g}(u_i,\bar u_i)|\Psi_{m_0}^N(\{u,\bar u_i\})\rangle
\een
with
 \ben
&& \label{Lamg}\Lambda_g^N(u,\bar u)=
\chi
 c(u)c(q^{-1}u^{-1}) \Lambda(u)\Lambda(q^{-1}u^{-1})G(u,\bar u),\label{Lamg}\\
&& \label{Eg}E_g^N(u_i,\bar u_i) =-\chi
\frac{ c(u_i)c(q^{-1}u_i^{-1}) }{b(q u_i^2)}\Lambda(u_i)\Lambda(q^{-1}u_i^{-1})G(u_i,\bar u_i).\label{Eg}
\eea
and
\bea
&& \chi=-\kappa \tilde \kappa \tau  \tilde\tau
 \Big(\frac{ \kappa \tau}{\tilde \kappa  \tilde\tau}+\frac{\tilde \kappa  \tilde \tau}{ \kappa \tau}
 +\frac{\xi  \tilde \mu}{ \tilde \xi \mu} q^{N+1}+\frac{\tilde \xi \mu}{\xi \tilde \mu} q^{-N-1}\Big).
 \een
The function $\Lambda(u)$ is given by (\ref{defLAM}) and the constant factor $\hat\eta_{m_0}$ by (\ref{etahatdef}). The
auxiliary functions $c(u)$, $b(u)$, $\tilde{F}(u,v)$ and $G(u,v)$ are collected
in appendix \ref{App:Func}, see equations (\ref{importantdef1}) and (\ref{importantdef2}).
\end{prop}

The proof of this proposition will be postponed to the next section. 

\begin{rmk}
This new term gives the inhomogeneous part of the modified T-Q Baxter equation introduced in the ODBA approach \cite{CYSW3} and recovered from the SoV approach in \cite{KMN14}. 
\end{rmk}

\begin{rmk}
The expression of the off-shell action of the modified creation operators $ \mathscr{B}_{dl}$ conjectured in \cite{BelPim14} have additional terms. These terms come from the shift of $m_0$ by $-2$ of the l.h.s of the equation (\ref{offshellB}) here by comparison of the l.h.s of the equation (5.13) in \cite{BelPim14}. This difference is due to the way we apply the MABA: in  \cite{BelPim14}, we put the transfer matrix into a triangular modified form and here we  put the transfer matrix into a diagonal modified form. In both cases the off-shell Bethe vector are the same, up to an overall constant.
\end{rmk}

The proposition \ref{P1} allows us to write the main result
of this paper, {\it i.e.}, the expression of the off-shell action of the
transfer matrix with two general boundaries on the Bethe vector (\ref{Psi}),
\bea\label{tronBV}
&&t(u)|\Psi_{m_0}^N(\bar u)\rangle=\Lambda^N(u,\bar u)|\Psi_{m_0}^N(\bar u)\rangle+
\sum_{i=1}^N\tilde{F}(u,u_i)E^N(u_i,\bar u_i)|\Psi_{m_0}^N(\{u,\bar u_i\})\rangle
\een
with
\ben\label{VP-BE}
&&\Lambda^N(u,\bar u)=\Lambda^N_{d}(u,\bar u)+\Lambda^N_g(u,\bar u),\quad E^N(u_i,\bar u_i) =E_{d}^N(u_i,\bar u_i)+E_g^N(u_i,\bar u_i),\label{LamE}
\een
where the modified diagonal part is given by (\ref{Lamdg},\ref{Edg}) and the off-diagonal contribution by (\ref{Lamg},\ref{Eg}).
For $E^N(u_i,\bar u_i) =0$, which are the Bethe's equations of the model, we obtain the on-shell action of the
transfer matrix which describes the eigenproblem of the transfer matrix (\ref{t}).


\section{proof of the off-shell action (\ref{offshellB}) \label{sectProof}}


In this section we prove the off-shell equation (\ref{offshellB}), which is the core equation of the MABA. We will 
use the left SoV basis of the dynamical $\mathscr{B}_{dl}$ operator constructed in \cite{FKN13}.
Let us consider the left-vector given by the product of the contravariant vectors (\ref{conVec}),
\ben
_{dl}\langle\Omega^N_{{m}}|=\otimes_{n=1}^N\,_{dl}\langle \tilde Y(v_n,m+n)|
\een
It is shown in \cite{FKN13}
that it is a pseudo-eigenvector of the dynamical $\mathscr{B}_{dl}$ operator\footnote{The term pseudo-eigenvector rather than eigenvector is due to the shift of $m$.}  , namely,
\ben
_{dl}\langle\Omega^N_{m}|\mathscr{B}_{dl}(u,m)=\,_{dl}\langle\Omega^N_{m-2}|\eta^N_m  \Lambda_b(u)
\een
with
\ben
\eta^N_m=-q \tilde \tau^2 \Big(q^{N+m} \beta_{dl}-i \frac{\tilde\mu\tau}{\mu\tilde\tau}\Big)\Big(q^{N+m} \beta_{dl}-i \frac{\mu\tau}{\tilde\mu\tilde\tau}\Big)\frac{\gamma^{dl}_m}{\gamma^{dl}_{m+N}\gamma^{dl}_{m+N+1}}
\een
and
\ben
 \Lambda_b(u)=u\,c(u) \prod_{n=1}^Nb(q u/v_n)b(u v_n).
\een
It follows that the vectors 
\ben\label{leftSoV}
\,_{dl}\langle\tilde \Psi_{m}(\bar h)|=
\,_{dl}\langle\Omega^N_{m}|\mathscr{A}^{h_1}_{dl}(v^{-1}_{1},{m+2})\dots\mathscr{A}^{h_N}_{dl}(v^{-1}_{N},{m+2})
\een
with $\bar h= \{h_1, \dots, h_N\}$ and $h_i\in \{0,1\}, \, i=1,...,N$, are pseudo-eigenvectors of the  dynamical $\mathscr{B}_{dl}$ operator,
\ben
\,_{dl}\langle\tilde \Psi_{m}(\bar h)|\mathscr{B}_{dl}(u,m)=
\left(\eta^N_m  \Lambda_b(u)\prod_{i=1}^N f(v_{i}^{-1},u)^{h_i}\right)\,_{dl}\langle\tilde \Psi_{m-2}(\bar h)|
\een
and its pseudo-spectrum is simple for generic inhomogeneity parameters, see \cite{FKN13} for details. Thus the SoV basis (\ref{leftSoV}) describe the full Hilbert space $\cH$.

\vspace{0.1cm}

Then, using the left-SoV-basis (\ref{leftSoV}), we can project the equation (\ref{offshellB}) on the $2^N$ one-dimensional subsets of the Hilbert space $\cH$. It reduces the equation (\ref{offshellB}) to functional relations which we can prove by usual analytical methods.
In fact, let us first consider the projection of the string of dynamical operators $\mathscr{B}_{dl}$ . From the properties of the SoV-basis we find,
\ben
&&\label{proj1}\,_{dl}\langle{\tilde \Psi}_{m_0+2(N-1)}
 (\bar h)|B_{dl}(\{\bar u,u\},m_0-2,N+1)|\Omega^N_{m_0}\rangle_{dr}=\\
&&\quad\qquad\Lambda_b(u) \Lambda_b(\bar u) W(\bar h)\prod_{i=1}^{N+1} \eta^N_{m_0+2(N-i)}\Big( \prod_{i=1}^N f(v_{i}^{-1},u)^{h_i}f(v_{i}^{-1},\bar u)^{h_i}\Big)\,_{dl}\langle\Omega^N_{{m_0-4}}|\Omega^N_{{m_0}}\rangle_{dr},\nonumber\\
&&\label{proj2}\,_{dl}\langle \tilde \Psi_{{m_0+2(N-1)}}(\bar h)|B_{dl}(\{\bar u_j,u\},m_0,N)|\Omega^N_{m_0}\rangle_{dr}=\\
&&\quad\qquad  \Lambda_b(u) \Lambda_b(\bar u_j)W(\bar h)\prod_{i=1}^N \eta^N_{m_0+2(N-i)}\Big( \prod_{i=1}^N f(v_{i}^{-1},u)^{h_i}f(v_{i}^{-1},\bar u_j)^{h_i}\Big) \,_{dl}\langle\Omega^N_{{m_0-2}}|\Omega^N_{{m_0}}\rangle_{dr},\nonumber
\een
where we have used the formula,
\ben\label{scalarinit}
&&\,_{dl}\langle \tilde \Psi_{{m}}(\bar h)|\Omega^N_{m_0}\rangle_{dr}=W(\bar h)\,_{dl}\langle\Omega^N_{{m}}|\Omega^N_{{m_0}}\rangle_{dr}, 
\een
with
\ben
&&W(\bar h)=\prod_{i=1}^N\Big(v_{i}^{-1}\, \tilde k^-(v_{i}^{-1}) \Lambda(v_{i}^{-1})\Big)^{h_i}
\een
which follows from the recursion relation\footnote{This recursion relation is proved in \cite{CYSW5}.},
\ben
&&\,_{dl}\langle\tilde \Psi_{{m}}(\bar h)|\Omega^N_{m_0}\rangle_{dr}=\Big(v_{N}^{-1}\, \tilde k^-(v_{N}^{-1}) \Lambda(v_{N}^{-1})\Big)^{h_N}\,_{dl}\langle
\tilde \Psi_{{m-2}}(\bar h_N)|\Omega^N_{m_0}\rangle_{dr}
\een
and where we have
\ben\label{scalarfinal}
&&\,_{dl}\langle\Omega^N_{{m}}|\Omega^N_{{m_0}}\rangle_{dr}=\prod_{n=1}^N\,_{dl}\langle \tilde Y(v_n,m+n)|X(v_n,m_0+n)\rangle_{dr}=\prod_{n=1}^N \frac{q \Big( \alpha_{dr} q^{-m_0-n} -\beta_{dl} q^{m+n} \Big)}{\gamma_{m+n+1}}.
\een

By means of these projections, we reduce the matrix equation (\ref{offshellB}) to the functional equations,
\ben\label{functionalsystem1}
&&-\chi u^{-1}c(qu)=
\frac{\Lambda^N_{g}(u,\bar u)}
{\Lambda_b(u)\prod_{j=1}^N f(v_j^{-1},u)^{h_j}}+
\sum_{i=1}^N
\tilde{F}(u,u_i)\frac{E^N_{g}(u_i,\bar u_i)}
{\Lambda_b(u_i)\prod_{j=1}^N f(v_j^{-1},u_i)^{h_j}},
\een
where we use
\ben
&&\hat \eta_{m_0}\eta^N_{m_0-2} \frac{\,_{dl}\langle\Omega^N_{{m_0-4}}|\Omega^N_{{m_0}}\rangle_{dr}}{\,_{dl}\langle\Omega^N_{{m_0-2}}|\Omega^N_{{m_0}}\rangle_{dr}}=-\chi.
\een

Using the explicit form (\ref{Lamg},\ref{Eg}), we reduce the proof to the checking the $2^N$ functional relations,
\ben\label{offshellBV4}
&&\sum_{i=1}^N\frac{1}{b(quu_i)b(u/u_i)}\frac{P(u_i,\bar v,\bar h)}{\prod_{n\neq i} b(qu_iu_n)b(u_i/u_n)} =\frac{P(u,\bar v,\bar h)}{\prod_{n=1}^{N} b(quu_n)b(u/u_n)}-1
\een
with
\ben
P(u,\bar v,\bar h)=\prod_{i=1}^{N}\big(b(uv_{i})b(qu/v_{i})\big)^{h_i}\big(b(quv_{i})b(u/v_{i})\big)^{1-h_i}.
\een
It can be proved by considering left and right side of the equation (\ref{offshellBV4}) as two meromorphic functions of $u$ and showing that they have the same poles and residues (at  $u=u_i$ and $u=q^{-1}u_i^{-1}$) on the full complex plane and the same values at $u=\infty$ and $u=0$. One uses the property $P(u,\bar v,\bar h)=P(q^{-1}u^{-1},\bar v,\bar h)$.

\section{Connection with the ODBA and the SoV methods  \label{sectODBA-SoV}}

 The ODBA and the SoV methods use the analytic properties of the transfer matrix. It is a Laurent series in $u$ of degree $2N+4$ and satisfy the parity condition and the crossing symmetry, see {\it e. g.} \cite{CYSW3}, 
\ben\label{parityt}
t(-u)=t(u), \quad t(q^{-1}u^{-1})=t(u).
\een
It follows from these two properties that the transfer matrix can be rewritten as a polynomial of 
\ben
U(u)=\frac{q u^2+q^{-1} u^{-2}}{q+q^{-1}}
\een
of degree $N+2$ with coefficients $t_i$ acting on $\cH$. 
\ben
t(u)=\sum_{i=0}^{N+2}t_i U(u)^i.
\een
We note that this variable $U(u)$ is used to extract the q-Onsager algebra from the reflection equation \cite{BK,BS}.
To fix any polynomial of degree $N+2$ one has to find $N+3$ values at distinct points. From the explicit form of the transfer matrix (\ref{tr}) in terms of the R-matrix (\ref{R}), the K-matrices (\ref{Kp}) and (\ref{Km}), it is a simple exercise to find that  
\ben\label{t1}
&&t(1)=(q+q^{-1})(\nu^++\nu^-)(\epsilon^++\epsilon^-)\prod_{i=1}^Nb(qv_i)b(qv_i^{-1}),
\een
\ben\label{ti}
&&t(i)=(q+q^{-1})(\nu^--\nu^+)(\epsilon^--\epsilon^+)\prod_{j=1}^Nb(iqv_j)b(iqv_j^{-1}),
\een
\ben\label{tinf}
&&\lim_{u\to \infty}t(u)=\frac{u^{4+2N}q^{2+N}}{(q-q^{-1})^{2N}}(\tilde \kappa^2 \tilde \tau^2+\kappa^2 \tau^2)+O(u^{2+2N})
\een
which fixes $3$ points. To get the remaining $N$ points the properties of the R-matrix and  the K-matrices, see \cite{CYSW3} for the proof, allows to relate the transfer matrix with the Sklyanin determinant from the relation
\ben\label{qdetKK} 
t(q^{-1}v_i)t(v_i)=\frac{{\rm Det}_q(K^+(q^{-1}v_i)K(q^{-1}v_i))}{b(q v_i^2)b(q v_i^{-2})}
\een
where
\ben
\qquad {\rm Det}_q(K^+(u)K(u))=b(u^2)b(q^{-4}u^{-2})\tilde k^+(qu)\tilde k^-(qu)\tilde k^+(q^{-1}u^{-1})\tilde k^-(q^{-1}u^{-1})\Lambda(qu)\Lambda(q^{-1}u^{-1}).
\een 
This relation is the core of the SoV and the ODBA methods. 
Using the polynomial properties of the transfer matrix, the $t(q^{-1}v_i)$ can be related to the $t(v_i)$ by Lagrange interpolation (see below the SoV characterization of the spectrum). It leads to $N$ quadratic relations on the $t(v_i)$ and gives at most $2^N$ solutions from the B\'ezout theorem. The ODBA proposes an analytical ansatz for the Baxter T-Q relation that satisfy the conditions (\ref{parityt}), (\ref{t1}), (\ref{ti}), (\ref{tinf}) and (\ref{qdetKK}) evaluated on some unknown eigenstate of the transfer matrix. 
 
\subsection{Connection with the ODBA} The key step in the ODBA was to propose a modified (or inhomogeneous) Baxter T-Q equation of the form
\ben\label{mTQ}
&&\Lambda^N(u,\bar u)Q^N(u,\bar u)=\psi(u)Q^N(q^{-1}u,\bar u)+\psi(q^{-1}u^{-1})Q^N(q u,\bar u)
 +\chi
 c(u)c(q^{-1}u^{-1}) \Lambda(u)\Lambda(q^{-1}u^{-1})
\een
with 
\ben
Q^N(u)=G(u,\bar u)^{-1},\quad 
\psi(u)=\phi(u)\tilde k^+(u)\tilde k^-(u)\Lambda(u)
\een
and the Bethe equation (\ref{VP-BE}) ensuring the analyticity of the eigenvalues. 
The most important point is the addition of an inhomogenous term in the ansatz such that the conditions (\ref{parityt}), (\ref{t1}), (\ref{ti}), (\ref{tinf}) and (\ref{qdetKK}) evaluated on some unknown eigenstate of the transfer matrix are satisfied. In the MABA context, the relation (\ref{mTQ})
appears naturally as a consequence of the algebraic properties of the double-row monodromy operators, see (\ref{VP-BE}).

\vspace{0.1cm}

In \cite{CYSW3} another ansatz depending on the parity of the length of the chain $N$ was proposed, actually it appears that there is an infinite family of modified Baxter T-Q equations which parametrize the spectrum of the transfer matrix. Here, from the MABA we recover the minimal solution with $N$ Bethe roots (\ref{mTQ}) that is equivalent of (\ref{LamE}). This minimal modified Baxter T-Q equation and the associated Bethe equations were shown from the SoV to give the complete description of the spectrum for the inhomogeneous case \cite{KMN14} and numerical evidence show that it should remains the case in the homogeneous limit.

\vspace{0.1cm}

Moreover, from the modified Baxter T-Q equation (\ref{mTQ}) and projection of the Bethe vector with the left SoV basis similar to the one we use here, the on-shell Bethe vector has been obtained in \cite{CYSW4,CYSW5}. The MABA gives the algebraic construction of the off-shell Bethe vector and allows, by limit or by imposing constraints between left and right boundary parameters, to recover the previous results in the literature, see \cite{BelPim14} for details.   

\subsection{Connection with the SoV characterization of the spectrum}

The SoV representation of the spectral problem of the transfer matrix is given \cite{FKN13}  (up to some restrictions for the values of the boundary parameters and of the inhomogeneous parameters, see in \cite{FKN13}) by
\ben\label{inter}
\Lambda^N(u)=\sum_{j=1}^N \Lambda^N(v_j) g_j(u)+f(u)
\een
with
\ben
g_j(u)=\frac{U(u)^2-1}{U(v_j)^2-1}\prod_{k=1, k\neq j}^N\frac{U(u)-U(v_k)}{U(v_j)-U(v_k)}
\een
and
\ben
f(u)=\Lambda^N(1)\frac{U(u)+1}{U(1)+1}\prod_{k=1}^N\frac{U(u)-U(v_k)}{1-U(v_k)}
+\Lambda^N(i)\frac{U(u)-1}{U(i)-1}\prod_{k=1}^N\frac{U(v_k)-U(u)}{1+U(v_k)}\\
+ \frac{(q+q^{-1})^{N+2}}{(q-q^{-1})^{2N}}(\tilde \kappa^2 \tilde \tau^2+ \kappa^2 \tau^2)(U(u)^2-1)\prod_{k=1}^N(U(u)-U(v_k)),\qquad\qquad\nonumber
\een
provided that the $\Lambda^N(v_j)$ satisfy the quadratic relations induced by the relation (\ref{qdetKK}) and the interpolation formula (\ref{inter}) at $u=q^{-1}v_i$. It is proven in \cite{KMN14} that this representation is complete and that it is equivalent to the modified T-Q Baxter equation (\ref{mTQ}).

\subsection{Connection with the SoV states}

To give the SoV states, let us recall the right-SoV-basis used to construct the eigenstates of the transfer matrix. Following \cite{FKN13} we introduce the right pseudo-vacuum 
\ben
|\Omega^N_{{m}}\rangle_{dl}=\otimes_{n=1}^N| Y(v_n,m+n)\rangle_{dl}
\een
that is a pseudo-eigenstate of the $\mathscr{B}_{dl}$ operator 
\ben
\mathscr{B}_{dl}(u,m)|\Omega^N_{{m}}\rangle_{dl}=\tilde \eta^N_m\tilde \Lambda_b(u)|\Omega^N_{{m+2}}\rangle_{dl}
\een
with
\ben
\tilde \eta^N_m=-q \tilde \tau^2 \Big(q^{-N+m} \beta_{dl}-i \frac{\tilde\mu\tau}{\mu\tilde\tau}\Big)\Big(q^{-N+m} \beta_{dl}-i \frac{\mu\tau}{\tilde\mu\tilde\tau}\Big)\frac{1}{\gamma^{dl}_{m+1}}
\een
and
\ben
\tilde \Lambda_b(u)=u\,c(u) \prod_{n=1}^Nb(q u v_n)b(u/v_n).
\een
The other states are constructed from the action of the operator
\ben
\mathscr{\hat D}(u,m)=\frac{\gamma_{m+1}}{\gamma_m}\langle\tilde X(u,m+2)|K(u)|Y(u^{-1},m)\rangle
\een
and are given by
\ben
|\tilde \Psi_{m}(\bar h)\rangle_{dl}=\prod_{i=1}^N\Big( \mathscr{\hat D}_{dl}(v_i,m)\Big)^{1-h_i}|\Omega^N_{{m}}\rangle_{dl}.
\een
The action of the $\mathscr{B}_{dl}(u,m)$ operator on these states is
\ben
\mathscr{B}_{dl}(u,m)|\tilde \Psi_{m}(\bar h)\rangle_{dl}=\tilde \eta^N_m\tilde \Lambda_b(u) \prod_{i=1}^N\Big(f(v_i,u)\Big)^{1-h_i}|\tilde \Psi_{m+2}(\bar h)\rangle\,_{dl}.
\een

The SoV states are given in \cite{FKN13} , up to an overall normalization, by  
\ben
|\Phi\rangle_{SoV}=\sum_{\bar h}\frac{1}{\mu^N_m(\bar h)} \prod_{j=1}^N \Big( \frac{ \Lambda^N(v_j^{-1}) }{v_j\tilde k^+(v_j^{-1})\phi(v_j^{-1})}\Big)^{h_j} |\tilde \Psi_{m+2}(\bar h)\rangle_{dl},
\een
where the $ \Lambda^N(v_j^{-1})$ are fixed from the quadratic relation (\ref{qdetKK}) and the interpolation formula (\ref{inter}).
We note that in \cite{FKN13} the gauge parameter $\beta$ is arbitrary and we restrict here to the case $\beta=\beta_{dl}$ to make the link with the Bethe vector  (\ref{Psi}) with $M=N$.
The $\mu^N_m(\bar h)$ is the diagonal element of the scalar product between left and right SoV basis 
\ben\label{ortho}
_{dl}\langle\tilde \Psi_{m}(\bar h)|\tilde \Psi_{m+2}(\bar k)\rangle_{dl}=\mu^N_m(\bar h)\prod_{i=1}^N\delta_{h_i, k_i} 
\een
explicitly given by
\ben
&&\mu^N_m(\bar h)=
\prod_{i=1}^N 
\frac{\bar \eta^N_{m-2(i-1)}}{\gamma^{dl}_{m+1+i}} v_i^{-2h_i+1} c(v_i) b(q^{-2h_i+1}  v_i^2)\nonumber\\&&\times
\prod_{j<i}^N
b(q^{-2h_j+1}v_j/v_i)
b(q^{-h_j-h_i+1}v_i/v_j)
b(q^{-2h_j+1}v_i v_j)
b(q^{h_j-h_i}v_i v_j)
\een
where
\ben
\bar \eta^N_{m}=\eta^N_{m} \frac{\gamma^{dl}_{m+N}\gamma^{dl}_{m+1+N}}{\gamma^{dl}_{m}}.
\een
To relate this result with the on-shell Bethe vector (\ref{Psi}) with $M=N$  constructed from the MABA we project both vectors with the left SoV-basis.  For the Bethe vector we use the formula (\ref{proj2})  with $u=u_i$ and for the SoV vector we use the scalar product between left and right SoV-basis (\ref{ortho}).
Thus we obtain in both cases, up to a overall constant, the same function, for each of the $2^N$ components, that show the equivalence of the two characterizations for the eigenstates of the transfer matrix.   

\section{conclusion \label{sectCon}}

In this paper we have proved the expression of the off-shell action of the modified creation operator on the Bethe vector, see proposition \ref{P1}. Thus we give the proof of the off-shell action of the transfer matrix of the XXZ spin chain on the segment conjectured in \cite{BelPim14}  and before in some limit cases \cite{BC13,Bel14}. The proof uses the left SoV basis \cite{FKN13} and exhibits similarities with the proof of the on-shell Bethe vector from the ODBA method \cite{CYSW5}. 

\vspace{0.1cm}

This shows the deep relation between the MABA, the ODBA and the SoV methods. Indeed, the MABA explains the algebraic origin of the inhomogenous term in the modified T-Q Baxter equation introduced in \cite{CYSW3} from the ODBA, as a modification of the analytical Bethe ansatz. Moreover it provides a correspondence between the Bethe vector and the eigenvectors in the SoV framework, completing the link already done for the eigenvalues in \cite{KMN14}. Let us add that it implies the existence of a modified coordinate Bethe ansatz whose wave function can be obtained by projection of the on-shell Bethe vector on the appropriate basis, extending the constrained boundary cases obtained in \cite{CRS1,CRS2} to generic boundaries. Thus the MABA incorporates the different BA (ODBA, Analytical BA, ABA, Coordinate BA) and is equivalent in the inhomogeous case to the SoV. 

\vspace{0.1cm}

The SoV method yields a proof of the completeness of the spectrum, at least in the inhomogenous case, which  can be extended to the MABA and, by complementarity, the MABA provides a regularization for the SoV such that the homogenous limit is well defined. We remark that such a connection was already used in the case of the XXX spin chain on the circle in \cite{MTV}, where the completeness of the Bethe ansatz solution was proved. Thus it seems that the complete description of the spectral problem of finite dimensional quantum integrable models constructed from the Yang-Baxter and reflection equations can be handled by both the Bethe ansatz and SoV methods. 

\vspace{0.1cm}

It could be also of interest to relate the MABA for the XXZ spin chain on the segment with other known methods such as the q-Onsager approach \cite{BK} or the non-polynomial solution from the homogeneous Baxter T-Q relation \cite{LP14}.

\vspace{0.1cm}

The knowledge of the Bethe vector is the first step to consider the correlation functions of the  XXZ spin chain on the segment in the ABA framework. An important question will be to understand the structure of the scalar product of the Bethe vector and see if some determinant representation can be obtained by analogy with the diagonal case \cite{KKMNST}. Let us mention the recent algebraic functional approach \cite{Gal14}  used in the diagonal case that should allow to find integral representations of the scalar product from our off-shell results. 

\vspace{0.1cm}

Other interesting questions will be to apply the MABA to other models such as periodic spin chains without $U(1)$ symmetry where the ODBA \cite{CYSW1,CYSW4} and the SoV \cite{SklySoV2,niccolisl2} methods were applied as well as developing a Nested MABA for spin chains without $U(1)$ related to higher rank algebras where only the spectrum was obtained in few cases from the ODBA, see {\it e. g.} \cite{CYSW6}.    

\vspace{0.2cm}

{\bf Acknowledgements:} S.B. and R.A.P. thank R. Nepomechie for discussions. S.B. is supported by the Universit\'e de Cergy-Pontoise post doctoral fellowship. S.B. is also partially supported by Sao Paulo Research Foundation (FAPESP), grant \# 2014/09832-1
and thanks the Departamento de F\'{\i}sica of the Universidade Federal de S\~ao Carlos for hospitality where a part of this work was done. R.A.P. is supported by FAPESP, grants \# 2014/00453-8 and \# 2014/20364-0. N.G. is supported by the Universit\'e de Cergy-Pontoise.

\appendix

\section{Functions\label{App:Func}}
We use the following functions throughout the text,
\ben
&&\label{importantdef1}b(u)=\frac{u-u^{-1}}{q-q^{-1}},
\quad k^-(u) =\nu_-u+\nu_+u^{-1},
\quad k^+(u) =\epsilon_+u+\epsilon_-u^{-1},\quad  c(u)=u^2-u^{-2},\\
&&\label{importantdef2}\phi(u)= \frac{b(q^2u^2)}{b(qu^2)},
\quad G(u,v)=\frac{1}{b(u/v)b(quv)},  \quad \tilde F(u,v)=(v/u) G(u,v)\frac{b(q^2 u^2)}{\phi(v)}, \\
&&f(u,v)= \frac{b(qv/u)b(uv)}{b(v/u)b(quv)}\ ,
\quad g(u,v)= \frac{\phi(q^{-1}v^{-1})}{b(u/v)}, \quad w(u,v)= -\frac{1}{b(quv)},\\
 &&h(u,v)= \frac{b(q^2uv)b(qu/v)}{b(quv)b(u/v)},\quad k(u,v)= \frac{\phi(u)}{b(v/u)}, \quad
n(u,v)= \frac{\phi(u)\phi(q^{-1}v^{-1})}{b(quv)},
\een
\ben
&& \gamma(u,m)=\alpha q^{-m} u-\beta q^{m} u^{-1}, \quad  \gamma(1,m)=\gamma_m,\\
&&g(u,v,m)=\frac{\gamma(u/v,m+1)}{\gamma_{m+1}}g(u,v), \quad w(u,v,m)=\frac{\gamma(uv,m)}{\gamma_{m+1}}w(u,v),\\
 &&k(u,v,m)=\frac{ \gamma(v/u,m+1)}{\gamma_{m+1}}k(u,v), \quad
n(u,v,m)=\frac{\gamma(1/(uv),m+2)}{\gamma_{m+1}} n(u,v),
\een

The following functional relations hold 
\ben\label{FR1}
f(u,\bar u)+\sum_{i=1}^M g(u,u_i,m-2)f(u_i,\bar u_i)-\phi(u_i)w(u,u_i,m-2)h(u_i,\bar u_i)=\frac{\gamma_{m-2M-1}}{\gamma_{m-1}},
\een
and
\ben\label{FR2}
h(u,\bar u)+\phi(u)^{-1}\sum_{i=1}^M\phi(u_i)k(u,u_i,m-2)h(u_i,\bar u_i)-n(u,u_i,m-2)f(u_i,\bar u_i)=\frac{\gamma_{m-2M-1}}{\gamma_{m-1}}.
\een

To prove these functional relations we consider them as a function of $u$. One can easily check that the left hand side of these equations are holomorphic functions on the full complex plane, thus they are constant. To evaluate these constants one sends $u$ to infinity and use the relation  
\ben\label{FR3}
\sum_{i=1}^M\phi(u_i)h(u_i,\bar u_i)+\phi(q^{-1}u^{-1}_i)f(u_i,\bar u_i)=\frac{q^{2M}-q^{-2M}}{q-q^{-1}}.
\een
This relation can be proved in a similar way. The left hand side is a symmetric function of the set $\bar u$, so one can consider it as a single variable function of any $u_k$ and show that for generic parameter $\bar u_k$ this function is holomorphic on the full complex plane, thus it is constant. Then sending $u_k$ to infinity one find a recursion relation on $M$ that fix the constant.



\begin{thebibliography}{99}


\bibitem{baxter}
Baxter R.J.,  \textsl{Exactly solved models in statistical mechanics} Academic 
Press, (1982).
%

\bibitem{BK} 
Baseilhac P. and Koizumi K.,
{\it Exact spectrum of the XXZ open spin chain from the q-Onsager algebra representation theory}, 
J. Stat. Mech. (2007) P09006, \texttt{arXiv:hep-th/0703106}

\bibitem{BS} 
Baseilhac P. and Shigechi K.,
{\it A new current algebra and the reflection equation},
Lett. Math. Phys. 92 (2010) 47-65, 
\texttt{arXiv:0906.1482}

\bibitem{Bel14}
Belliard S., 
{\it Modified algebraic Bethe ansatz for XXZ chain on the segment - I - triangular cases}, 
Nucl. Phys. \textbf{B892} (2015) 1-20, \texttt{arXiv:1408.4840}

\bibitem{BC13} 
Belliard S.  and Cramp\'e N., 
\textsl{Heisenberg XXX model with general boundaries: Eigenvectors from Algebraic Bethe ansatz},
SIGMA {\bf 9}  (2013) 072, \texttt{arXiv:1309.6165} 


\bibitem{BCR12} 
Belliard S., Cramp\'e N. and Ragoucy E., 
{\it Algebraic Bethe ansatz for open XXX model with triangular boundary matrices}, 
Lett. Math. Phys. {\bf 103} (2013) 493, \texttt{arXiv:1209.4269}


\bibitem{BelPim14}
Belliard S. and Pimenta R.A.,
{\it Modified algebraic Bethe ansatz for XXZ chain on the segment - II - general cases}, 
(2014), Nucl. Phys. \textbf{B894} (2015) 527-552, \texttt{arXiv:1412.7511}


\bibitem{gauge}
Cao J. , Lin H.-Q. , Shi K. and Wang Y.,
\textit{Exact solutions and elementary excitations in the XXZ spin chain with unparallel boundary fields},
Nucl. Phys. \textbf{B663} (2003) 487, \texttt{arXiv:cond-mat/0212163}

\bibitem{CYSW1} 
Cao J., Yang W., Shi K. and Wang Y.,
 {\it Off-diagonal Bethe ansatz and exact solution of a topological spin ring},
\textsl{Phys. Rev. Lett.} \textbf{111} (2013) 137201,  \texttt{arXiv:1305.7328}

\bibitem{CYSW3}
Cao J., Yang W., Shi K. and Wang Y.,
{\it Off-diagonal Bethe ansatz solutions of the anisotropic spin-1/2 chains with arbitrary boundary fields}, Nucl. Phys. {\bf B877} (2013) 152-175,
\texttt{arXiv:1307.2023},

Cao J., Yang W., Shi K. and Wang Y., 
{\it On the complete-spectrum characterization of quantum integrable spin chains via the inhomogeneous T-Q relation}, \texttt{arXiv:1409.5303}

\bibitem{CYSW6} 
Cao J., Yang W., Shi K. and Wang Y., 
{\it Nested off-diagonal Bethe ansatz and exact solutions of the su(n) spin chain with generic integrable boundaries},  JHEP {\bf 04} (2014) 143, {\it arXiv:1312.4770}

\bibitem{Cra14} 
Cramp\'e N., \textsl{Algebraic Bethe ansatz for the totally asymmetric simple exclusion process with boundaries}, J. Phys. A: Math. Theor. {\bf 48} (2015) 08FT01, \texttt{arXiv:1411.7954}


\bibitem{CRS1} 
Cramp\'e N., Ragoucy E. and Simon D.,
{ \it Eigenvectors of open XXZ and ASEP models for a class of non-diagonal boundary conditions},
J. Stat. Mech. (2010) P11038, \texttt{arXiv:1009.4119}

\bibitem{CRS2} 
Cramp\'e N., Ragoucy E.  and Simon D.,
{\it Matrix Coordinate Bethe Ansatz: Applications to XXZ and ASEP models},
\textsl{J. Phys.} \textbf{A44} (2011) 405003, \texttt{arXiv:1106.4712}


\bibitem{dVGR1} 
de Vega H. J. and Gonzalez-Ruiz A.,
\textsl{Boundary K-matrices for the six vertex and the n(2n-1) $A_{n-1}$ vertex models}, 
J. Phys. \textbf{A26} (1993) 519, \texttt{arXiv:hep-th/9211114}





\bibitem{FKN13} 
Faldella S., Kitanine N. and Niccoli G., 
{\it Complete spectrum and scalar products for open spin-1/2 XXZ quantum chains with non-diagonal boundary terms}, 
J. Stat. Mech. (2014) P01011, \texttt{arXiv:1307.3960}


\bibitem{Gal08} 
Galleas W.,
{ \it Functional relations from the Yang-Baxter algebra: Eigenvalues of the XXZ model with non-diagonal twisted and open boundary conditions}, 
Nucl. Phys. \textbf{B790} (2008) 524, \texttt{arXiv:0708.0009}

\bibitem{Gal14} 
Galleas W.,
{\it Off-shell scalar products for the $XXZ$ spin chain with open boundaries}
Nucl. Phys. B893, (2015) 346-375, \texttt{arXiv:1412.5389}




\bibitem{KKMNST}
Kitanine N., Kozlowski K.K., Maillet  J.M., Niccoli G., Slavnov N.A. and Terras V.,  
{\it Correlation functions of the open XXZ chain I}, 
J. Stat. Mech. (2007) P10009, {\tt arXiv:0707.1995}
 
Kitanine N., Kozlowski K.K., Maillet  J.M., Niccoli G., Slavnov N.A. and Terras V., 
{\it Correlation functions of the open XXZ chain II}, 
J. Stat. Mech. (2008) P07010, {\tt arXiv:0803.3305}

\bibitem{KMN14}
Kitanine N., Maillet J.M. and Niccoli G.,
{\it Open spin chains with generic integrable boundaries: Baxter equation and Bethe ansatz completeness from separation of variables}, 
J. Stat. Mech. (2014) P05015, {\tt arXiv:1401.4901}.


\bibitem{LP14}
Lazarescu A., Pasquier V., {\it Bethe Ansatz and Q-operator for the open ASEP}
J. Phys. A: Math. Theor. {\bf 47} (2014) 295202, {\tt  arXiv:1403.6963 }

\bibitem{MS96}
Maillet J.M. and Sanchez de Santos J.,{\it  Drinfel'd Twists and Algebraic Bethe Ansatz}, Amer. Math. Soc. Transl. 201 (2000) 137, {\tt arXiv:q-alg/9612012}.

\bibitem{MTV}
 Mukhin E., Tarasov V. and Varchenko A., 
 {\it Bethe algebra of homogeneous XXX Heisenberg model has simple spectrum}, 
 Commun. Math. Phys. {\bf 288} (2009) 1, {\tt arXiv:0706.0688}

\bibitem{TQ}
Nepomechie R.I.,
\textit{Solving the open XXZ spin chain with nondiagonal boundary terms at roots of unity},
Nucl. Phys. \textbf{B622} (2002) 615-632, \texttt{arXiv:hep-th/0110116};

Nepomechie R.I.,
\textit{Bethe Ansatz solution of the open XXZ chain with nondiagonal boundary terms},
J. Phys. \textbf{A37} (2004) 433, \texttt{arXiv:hep-th/0304092}

\bibitem{TQ2}
Nepomechie R. I. and Ravanini F., 
\textit{Completeness of the Bethe Ansatz solution of the open XXZ chain with nondiagonal boundary terms}, J. Phys. \textbf{A36} (2003), Addendum ibid A 37 (2004) 1945, \texttt{arXiv:hep-th/0307095}

\bibitem{niccoli2}
Niccoli G.,  
{\it Non-diagonal open spin-1/2 XXZ quantum chains by separation of variables: Complete spectrum and matrix elements of some quasi-local operators}, 
J. Stat. Mech.  (2012) P10025, \texttt{arXiv:1206.0646}.

\bibitem{niccolisl2}
Niccoli G.,
{\it Antiperiodic spin-1/2 XXZ quantum chains by separation of variables: Complete spectrum and form factors}, Nucl.Phys. {\bf B870}  (2013) 397-420, {\tt arXiv:1205.4537}

Niccoli G. and Terras V., 
{\it Antiperiodic XXZ chains with arbitrary spins: Complete eigenstate construction by functional equations in separation of variables}, (2014) {\tt arXiv:1411.6488}


\bibitem{PL13} 
Pimenta R.A. and Lima-Santos A.,
{\it Algebraic Bethe ansatz for the six vertex model with upper triangular $K$-matrices},
J. Phys. A: Math. Theor. {\bf 46} (2013) 455002, \texttt{arXiv:1308.4446}


\bibitem{SFT}
Sklyanin E.K., Takhtadzhyan L.A. and Faddeev  L.D., 
{\it The Quantum Inverse Problem Method. I}, 
Theor. Math. Phys. \textbf{40} (1979), 688.

\bibitem{Skl88}
Sklyanin E.K., 
\textit{Boundary conditions for integrable quantum systems},
J. Phys.  \textbf{A21} (1988) 2375

\bibitem{SklySoV2}
Sklyanin E.K., 
{\it Quantum inverse scattering method. selected topics}, 
In M.-L. Ge, editor, Quantum group and Quantum Integrable Systems. Nankai Lectures in Mathematical Physics, World Scientific, (1992), {\tt ArXiv:hep-th/9211111}


\bibitem{YanZ07}
Yang  W.-L. and Zhang Y.-Z., 
 {\it On the second reference state and complete eigenstates of the open {$XXZ$} chain},
JHEP {\bf 0704} (2007) 044, \texttt{arXiv:hep-th/0703222}



\bibitem{CYSW4}  
Zhang X., Li Y.Y., Cao J., Yang W.-L., Shi K., Wang Y.,
{\it Retrieve the Bethe states of quantum integrable models solved via off-diagonal Bethe ansatz},  J. Stat. Mech. (2015) P05014, \texttt{arXiv:1407.5294v3}


\bibitem{CYSW5}  
Zhang X., Li Y.Y., Cao J., Yang W.-L., Shi K., Wang Y.,
{\it  Bethe states of the XXZ spin-1/2 chain with arbitrary boundary fields}, Nucl. Phys. \textbf{B893} (2015) 70 - 88,
\texttt{arXiv:1412.6905}


\end{thebibliography}
\end{document}